 \documentclass{ws-ijmpe}
\usepackage[super,compress]{cite}
\usepackage{graphicx}
\usepackage{xcolor}
\usepackage{lineno}

\begin{document}
\markboth{Domenico Colella}{Upgrade of the ALICE experiment beyond LHC Run~3}

\catchline{}{}{}{}{}
%
%%%%%%%%%%%%%%%%%%%%%%%%%%%%%%%%%%%%%%%%%%%%%%%%%%%%%%%%%%%%%%%%%%%%

\title{Upgrade of the ALICE experiment beyond LHC Run~3}

\author{Domenico Colella}
\address{Politecnico di Bari and INFN sezione di Bari, Via E. Orabona 4\\
Bari, 70125, Italy\\
domenico.colella@ba.infn.it}
\address{on behalf of the ALICE Collaboration}

\maketitle

\begin{history}
\received{Day Month Year}
\revised{Day Month Year}
\end{history}

\begin{abstract}
The ALICE Collaboration completed the upgrade of the detector and is now commissioning for the beginning of the data taking during LHC Run~3.
In parallel, R$\&$D activities and simulation studies are being performed to define the future of the experiment beyond LHC Run~3. 
Two detector upgrades are foreseen for the next long shutdown (LS3). The first is the replacement of the three layers of the Inner Tracking System closest to the beam with a novel vertex detector consisting of curved wafer-scale ultra-thin silicon sensors arranged in perfectly cylindrical layers to improve impact parameter resolution and significantly extend the physics capability for the study of the heavy-flavour production and the low-mass dielectrons. The second upgrade for the LS3 is the addition of a Forward Calorimeter detector at large rapidity consisting of a Si-W electromagnetic calorimeter with pad and pixel readout, that will equip the experiment with unique capabilities to measure small-x gluon distributions via prompt photon production. 
A proposal of a next-generation heavy-ion experiment for LHC Run~5 is also in preparation and will be discussed. The aim is to perform novel measurements of the electromagnetic and hadronic probes of the QGP, such as the production of multiply-charmed baryons, which have so far been inaccessible, both because of detector performance and luminosity. The concept of the new apparatus foresees an extensive usage of thin silicon sensors for tracking and a modern particle identification system, combining a silicon-based time of flight detector, a RICH detector, an electromagnetic calorimeter and a muon system. 

\end{abstract}

%\ccode{PACS numbers: 25.75.−q, }

\tableofcontents

%_______________
\section{Introduction}	
ALICE (A Large Ion Collider Experiment) \cite{JINST} is a general-purpose heavy-ion experi\-ment at the CERN LHC. Its main goal is to study the physics properties of the quark-gluon plasma (QGP). 

During LHC Run~1 (2009-2013) and Run~2 (2015-2018), the analysis of  the data collected in heavy-ion collisions allowed the observation of hot hadronic matter at unprecedented values of temperatures, densities and vo\-lu\-mes. These studies confirmed the basic picture, emerged from the experimental investigation at lower energies, of a QGP as an almost inviscid liquid and has, for example, provided clear evidence of the importance of regeneration in the production of J/$\Psi$ mesons in the freeze-out of the QGP, as well as precision measurements of energy loss of light and heavy quarks, allowing to determine the relevant transport coefficients.

The study of the strongly-interacting state of matter in the second generation of LHC heavy-ion studies in LHC Run~3 (2022-2024) and Run~4 (2027-2030) will focus on rare processes such as the production of heavy-flavour particles, quarkonium states, real and virtual photons and heavy nuclear states \cite{ALICEupLoI}. The earlier methods of triggering will be limited for many of these measurements, particularly at low-$p_{\rm T}$. Therefore, the \mbox{ALICE} collaboration planned to upgrade the LHC Run~1/Run~2 detector by enhancing its low-momentum vertexing and tracking capability, allowing data taking at substantially higher rates and preserving the already remarkable particle identification capabilities. A second upgrade phase, to be completed before LHC Run~4, foresees a further upgrade of the inner tracking system and the introduction of a new forward calorimeter detector.

The ALICE Collaboration is defining its future beyond LHC Run~4 (from 2032 onward) proposing a completely new LHC experiment to be installed in place of the current ALICE detector\cite{ALICE3EoI}. It will be dedicated to the high-statistics study of the production of heavy flavour hadrons and of soft electromagnetic and hadronic radiation produced in high-energy proton-proton and nuclear collisions, opening a new window which will provide a better understanding of the initial temperature of the QGP produced in the collisions, new insights in the interplay between thermalisation of heavy flavour and hadron formation, and in the nature of chiral symmetry breaking, beyond what is planned for the next ten years.

%___________________________________
\section{ALICE~2: readiness for LHC Run~3}
The LHC long shutdown 2 (LS2, 2019-2021) is going to be completed soon and ALICE is finally preparing for data taking during LHC Run~3. 
Here the list of upgrades ALICE underwent during the LS2:
\begin{itemize}
\item Reduction of the beam-pipe radius from \mbox{29.8 mm} to \mbox{19.2 mm}. 
\item Installation of two new high-resolution, high-granularity, low material budget silicon trackers:
  \begin{itemize}
  \item Inner Tracking System (ITS~2) \cite{ITSUPTDR} in the central pseudo-rapidity. 
  \item Muon Forward Tracker (MFT) \cite{MFTTDR} covering forward pseudo-rapidity. 
  \end{itemize}
\item Replacement of the endcap wire chambers of the Time Projection Chamber by Gas Electron Multiplier (GEM) detectors and installation of new readout electronics allowing continuous readout \cite{TPCUPTDR}.
\item Replacement of  Resistive Plate Chambers of the Muon Identifier that shown ageing effects during LHC Run~2\cite{RDOUPTDR}.
\item Upgrades of the forward trigger detectors (Fast Interaction Trigger, FIT) \cite{RDOUPTDR}.
\item Upgrades of the readout electronics of the Transition Radiation Detector, Time-Of-Flight, Photon Spectrometer, Muon Spectrometer and Zero Degree Calorimeter for high rate operation \cite{RDOUPTDR}.
\item Upgrades of online and offline systems (O$^{2}$ project) \cite{O2TDR} in order to cope with the expected data volume.
\end{itemize}

\begin{figure}[tb]
\centerline{\includegraphics[width=12.5cm]{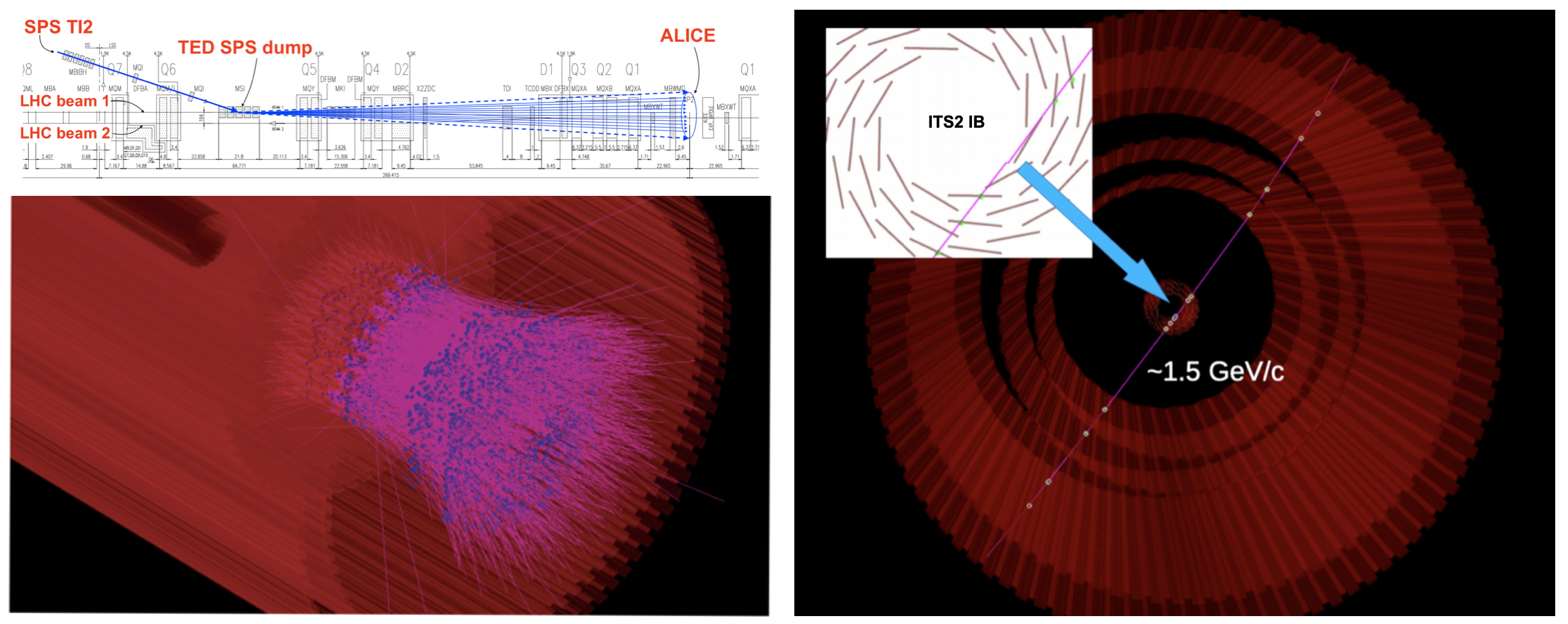}}
\caption{
Left: example of muon tracks, generated during the SPS to LHC transfer line test stopping the beam at the Target Extraction Dump (TED) close to P2, as reconstructed in the MFT detector. Right: example of cosmic ray track reconstructed in the ITS.}
\label{FigA}
\end{figure}

\begin{figure}[tb]
\centerline{\includegraphics[width=12.5cm]{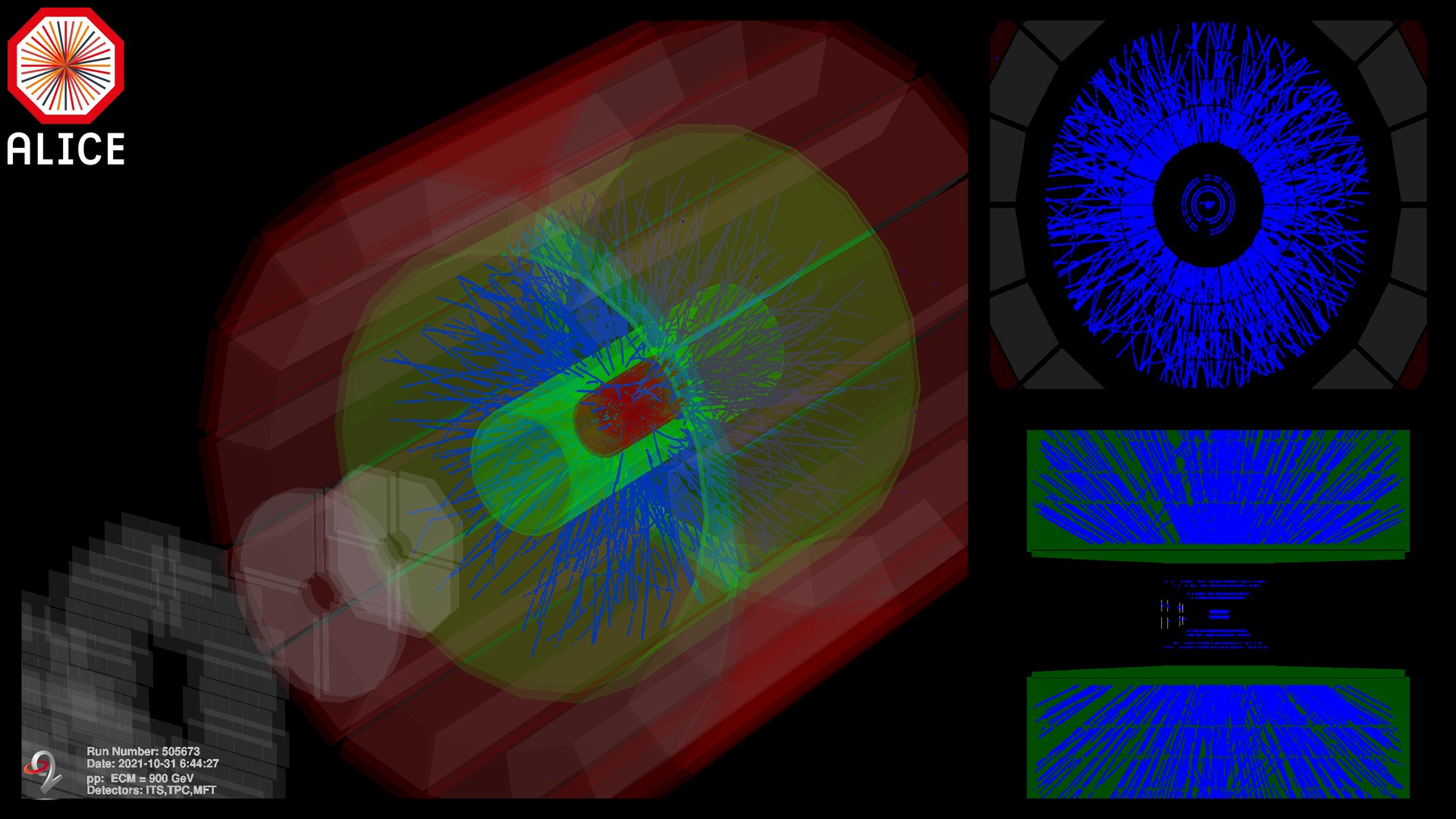}}
\caption{ALICE event display example for an event recorded during the proton-proton collisions provided by LHC during the pilot runs, 27 - 31 October 2021. Left: 3D view. Top right: transverse view. Bottom right: longitudinal view.}
\label{FigA3}
\end{figure}

All the needed interventions have been completed by August 2021, when \mbox{ALICE} started global commissioning with most of the detector systems. 
The effort in this period is focused on the integration of the single detectors into the common readout and detector control systems. During this period a large cosmic rays campaign has been carried out, in order to verify detector responses, data-acquisition workflow and detector control system. An example of cosmic ray track reconstructed in the ITS is shown in Fig. \ref{FigA} (right canvas). First global verification of detector performance with large amount of tracks happened at the beginning of October 2021 during the verification of the beam transfer line procedure between SPS and LHC. During these tests, the beam is dumped in the Target Extraction Dump, creating an intense flux of particles (mainly muons) at forward rapidity that could be seen by ALICE detector. Fig. \ref{FigA} (left canvas) shows the tracking capabilities of the MFT in this beam condition.
The following important commissioning step took place during the LHC pilot runs at the end of October 2021, when real proton--proton collisions have been provided by the machine allowing to perform first physics event reconstruction (Fig. \ref{FigA3}).

%_________________________________________
\section{ALICE~2: R$\&$D activities for LHC Run~4}
The physics program of the experiment during LHC Run~4 could be extended and improved thanks to recent innovations in the field of silicon imaging sensor technology that open extraordinary opportunities for new detector concepts. The ALICE Collaboration presented two Letters of Intent for a further upgrade of the Inner Tracking System (ITS~3)\cite{ITS3LoI} and for the installation of a calorimeter in the forward region (FoCal)\cite{FoCalLoI}.

\subsection{ITS~3}
The ITS~2 consists of seven cylindrical detector layers based on CMOS Monolithic Active Pixel Sensors (MAPS), named ALPIDE (ALice PIxel DEtector), covering a 10 m$^{2}$ area with about 12.5 billion pixels. This sensor has dimensions 3~$\times$~1.5~cm$^{2}$ and is thinned down to 50~$\mu$m (for the three innermost layers, inner barrels). Each layers is azimuthally segmented in units named staves, consisting of the following main components: a carbon fiber support structure, a carbon plate that embeds the cooling pipes, an assembly of ALPIDE sensors (whose number depends on the radial position of the corresponding layer) and a polyimide flexible printed circuit (for chip configuration and data streaming), and (for the four outermost layers) a power bus. The binary readout with zero suppression, implemented in the ALPIDE, allows the average power density to be below 40~mW/cm$^{2}$ hence granting operation at room temperature using water cooling.

\begin{figure}[tb]
\centerline{\includegraphics[width=12.5cm]{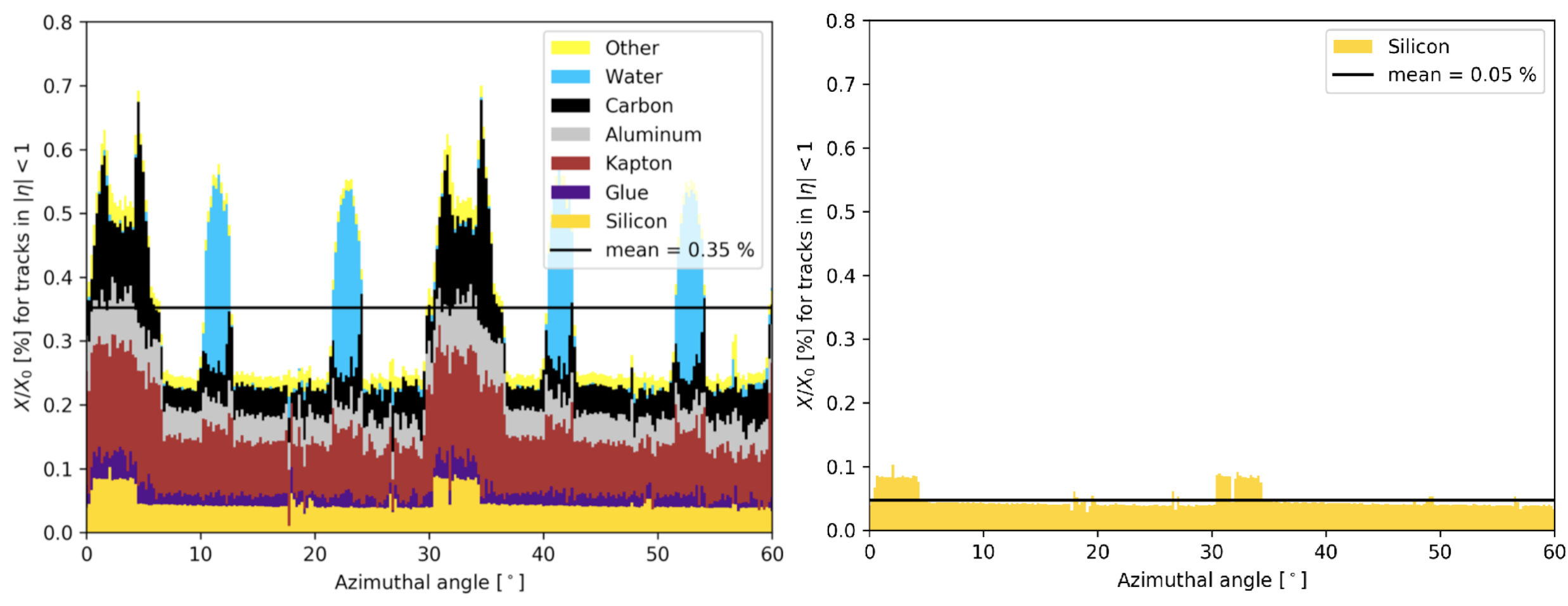}}
\caption{Material budget angular distribution of two adjacent staves in the ITS~2 configuration (left) and leaving only the silicon sensor (right). Thickness of the silicon is $\approx$ 0.05\%~X/X$_{0}$.}
\label{FigB}
\end{figure}

In parallel with the reduction of the radial position of the first layer, one of the key detector design feature to improve the momentum resolution and tracking performance, especially for particles with low transverse momentum (low-$p_{\rm T}$ $<$ 1~GeV/c), is the reduction of the material budget. ITS~2 design, with reduction of the support structures and the thickness of the sensor, brought to the excellent result of an estimated mean material budget value of 0.35\%~X/X$_{0}$ for each of the three innermost layers. The material budget angular distribution and the contribution from each component material, for two adjacent staves in the first layer of the ITS~2, is reported in Fig.~\ref{FigB} (left). The following considerations can be made:
\begin{itemize}
\item material budget is, to large extent, formed by passive components such as water cooling, carbon and kapton support structures, and aluminum wires;
\item a lot of irregularities, e.g. due to the overlap of adjacent staves (needed to grant hermeticity) or due to the presence of the water cooling pipes, are visible;
\item silicon makes up only about 15\% of the total material budget.
\end{itemize}

\begin{figure}[tb]
\centerline{\includegraphics[width=12.5cm]{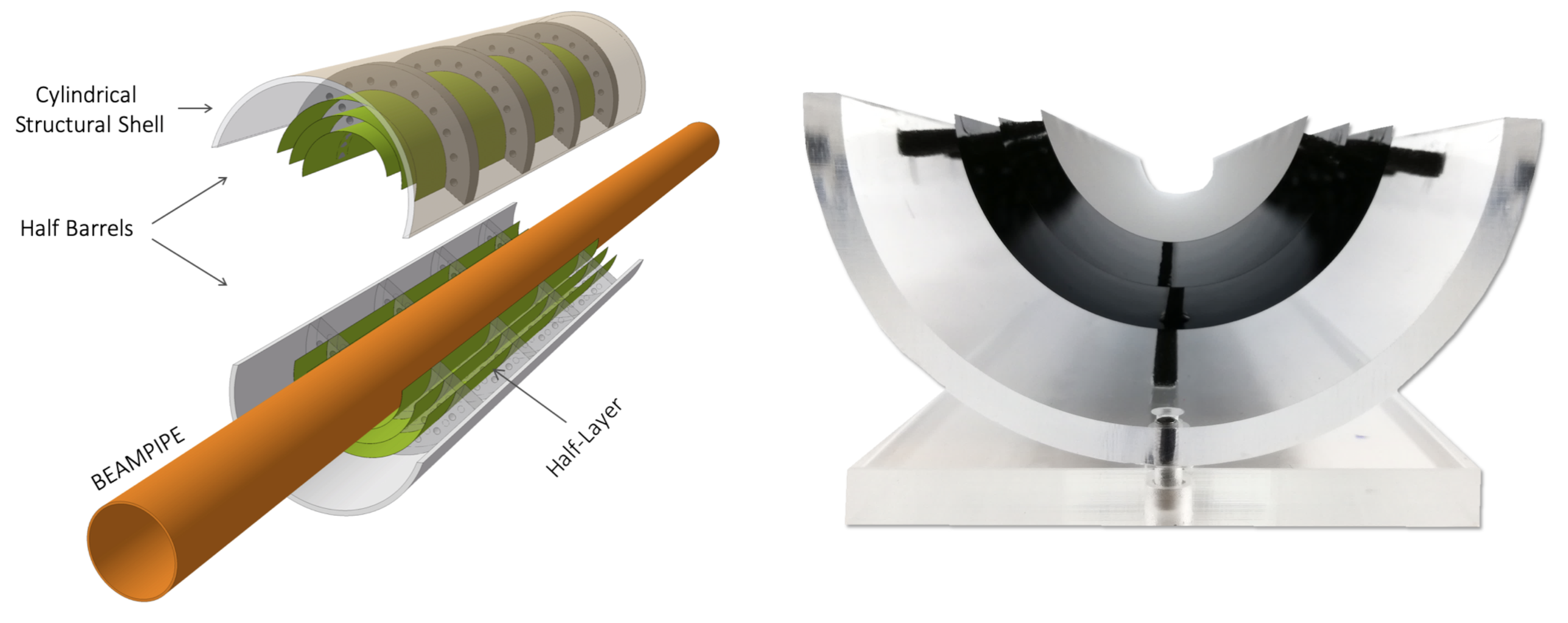}}
\caption{Left: proposed design for the inner barrel of the ITS in Run 4. Right: ITS 3 mechanical prototype assembly.}
\label{FigC}
\end{figure}

The new design of the ITS inner barrels for Run~4 aims to avoid the passive material keeping essentially just the silicon layer. Such a detector would achieve an unprecedented low material budget of about 0.05\%~X/X$_{0}$ per layer (Fig.~\ref{FigB}, right). The construction of the detector would require (i) to reduce the power consumption below 20~mW/cm$^{2}$, in order to cool down the sensor only by airflow, (ii) to integrate power and data buses on the chip, in order to remove any other flex covering the sensitive region, and (iii) to rely on the stiffness of large size, bent silicon wafers, in order to remove mechanical support structures. Such a low value for the power consumption can be achieved by moving the sensor periphery, including the serial link, to the edge of the chip and by the usage of the 65~nm CMOS technology. Large size MAPS with an area of up to 21~cm $\times$ 21~cm using wafers that are 300~mm in diameter, can be developed using a stitching technique. The reduction of the sensor thickness to values of about 20 -- 40~$\mu$m will open the possibility of exploiting the flexible nature of silicon to implement large-area curved sensors. In this way, it will become possible to build a cylindrical layer of silicon-only sensors, like depicted in Fig.~\ref{FigC} (left). Installation of a new beam pipe with smaller radius (inner radius 16~mm) and thickness (500 $\mu$m), will allow to place the first layer even closer to the interaction point (from 23~mm of the ITS~2 first layer to 18~mm). 

\begin{figure}[tb]
\centerline{\includegraphics[width=12.5cm]{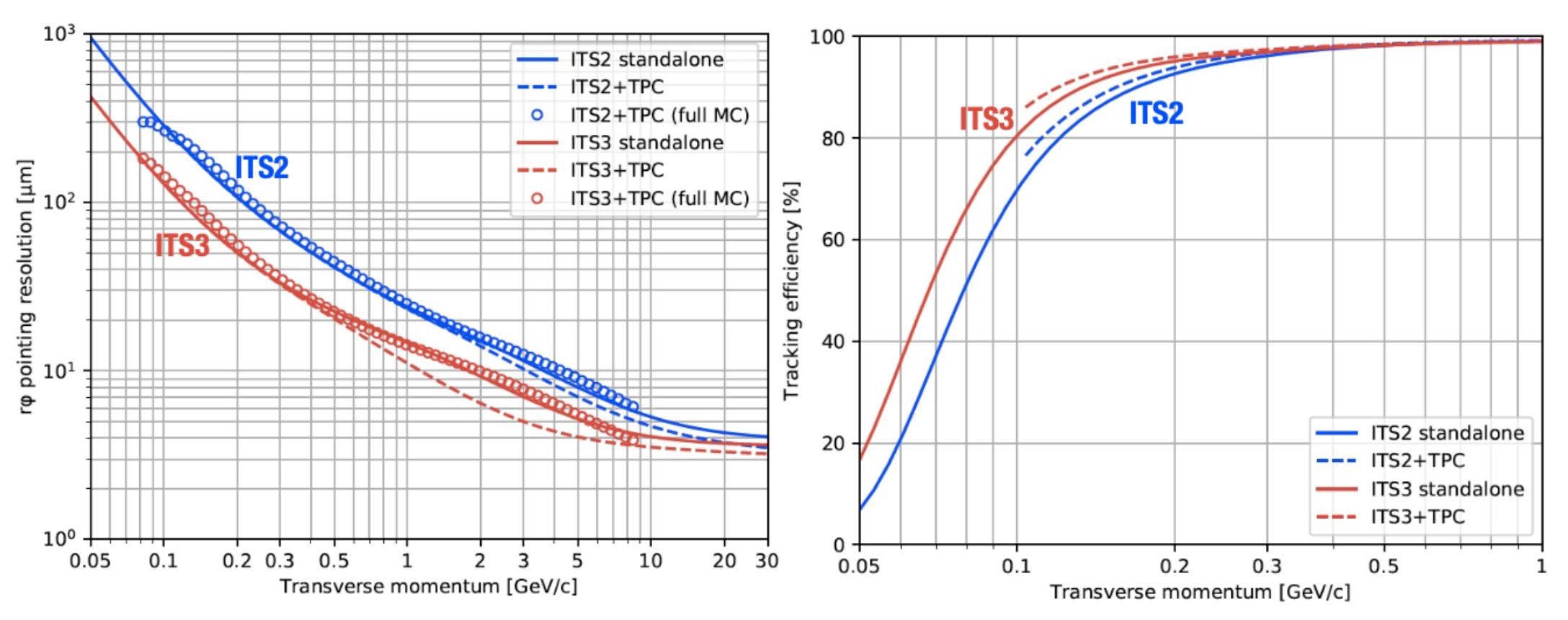}}
\caption{Comparison of the tracking performance in Run 3 (blue color) and Run 4 (red color). Left: pointing resolution of tracks in the plane transverse to the beam. Right: tracking efficiency.}
\label{FigD}
\end{figure}

Performance studies made for the ITS~3, indicate a further improvement of the pointing resolution by approximately a factor 2 and an increase of the stand-alone tracking efficiency by factor 2 for $p_{\rm T} <$~100~MeV/c (Fig. \ref{FigD}). The improvement of the vertexing performance and the reduction of material budget will have a dramatic impact on the measurement of charm and beauty hadrons at low transverse momentum as well as on the measurement of low-mass and low $p_{\rm T}$ dielectrons. The impact on various physics channels has been studied, showing for example that the significance of the $\Lambda_{c}$ in lead--lead collisions increases by almost a factor 4.

\begin{figure}[tb]
\centerline{\includegraphics[width=12.5cm]{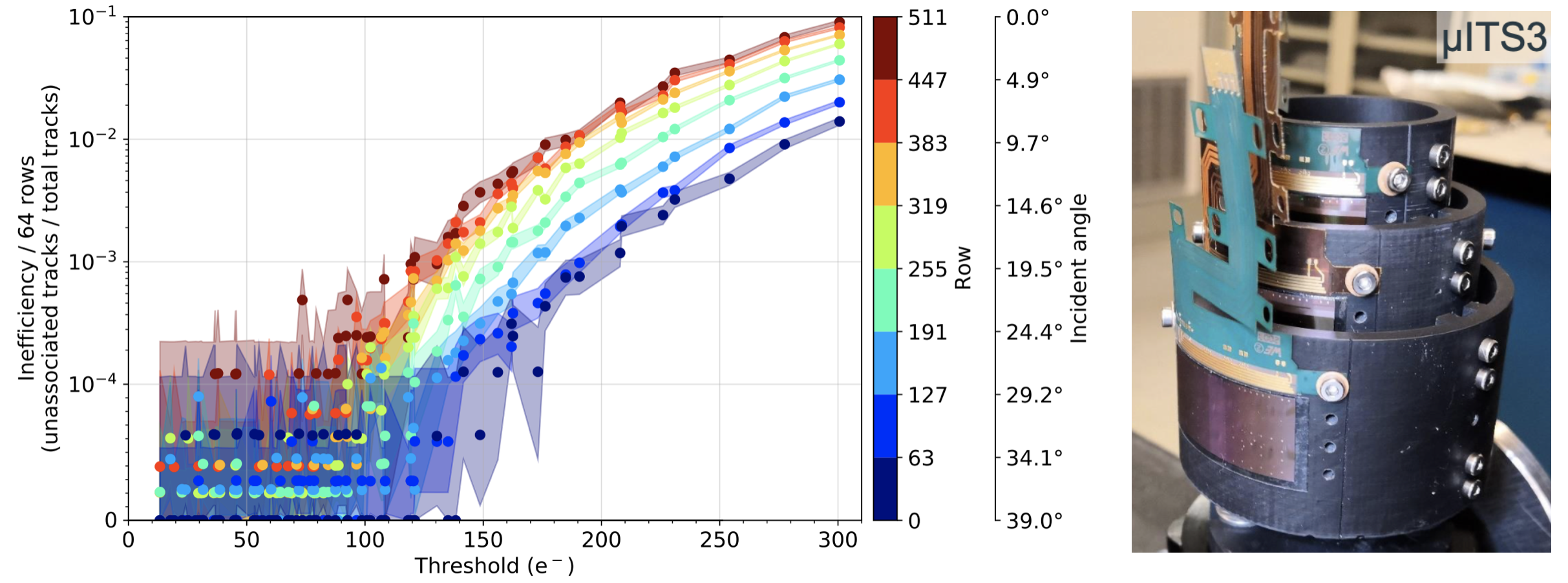}}
\caption{Left: hit inefficiency as a function of threshold for different rows and incident angles as measured in a bent ALPIDE. Right: $\mu$ITS3 assembly.}
\label{FigE}
\end{figure}

Current R\&D lines are exploring detector integration, sensor performance and chip design. 

The possibility to bend an ALPIDE chip has been explored, successfully reaching the ITS~3 target radii without breaking for a silicon thickness below 50~$\mu$m. An estensive study established the best commercially available carbon foam material to be used for mechanical supports, targeting an excellent thermal conductivity with a low density. A full size mechanical prototype, using large dimension, 50~$\mu$m thick, blank wafer, has been assembled, allowing the development of bending tools and verification of carbon foam support structures (Fig.~\ref{FigC}, right). 

The ALPIDE sensor characterised in the laboratory showed no effect on the performance due to the bending process; the noise level and number of dead pixels remained unchanged, and the difference in pixel threshold distribution over the matrix is negligible. The high detection efficiency of single ALPIDE sensor has been measured at the DESY test beam facility, and found to be preserved also in bent configuration (Fig.~\ref{FigE}, left): below the threshold of 100~e$^{-}$ (nominal operating point in ITS~2) the hit inefficiency is generally lower than 10$^{-4}$ independently of the incident angle or the position on the chip. Additionally, an assembly, made of 6 ALPIDE sensors bent around cylinders of ITS~3 target radii and having open windows in correspondence to the sensors (the $\mu$ITS3, Fig. \ref{FigE}, right), has been tested with beam to verify tracking and vertexing capabilities. 

New chip design in 65 nm reached first milestone in June 2021 with the production of the multi-layer reticle 1 (MLR1) including first test structures like: transistor test structures, analog building blocks, various diode matrices and digital test matrices. Characterisation of these structures in laboratory and under beam are ongoing. The next important milestone will be the first engineering run including first implementation of the stitching technique.

\subsection{FoCal}

\begin{figure}[tb]
\centerline{\includegraphics[width=8cm]{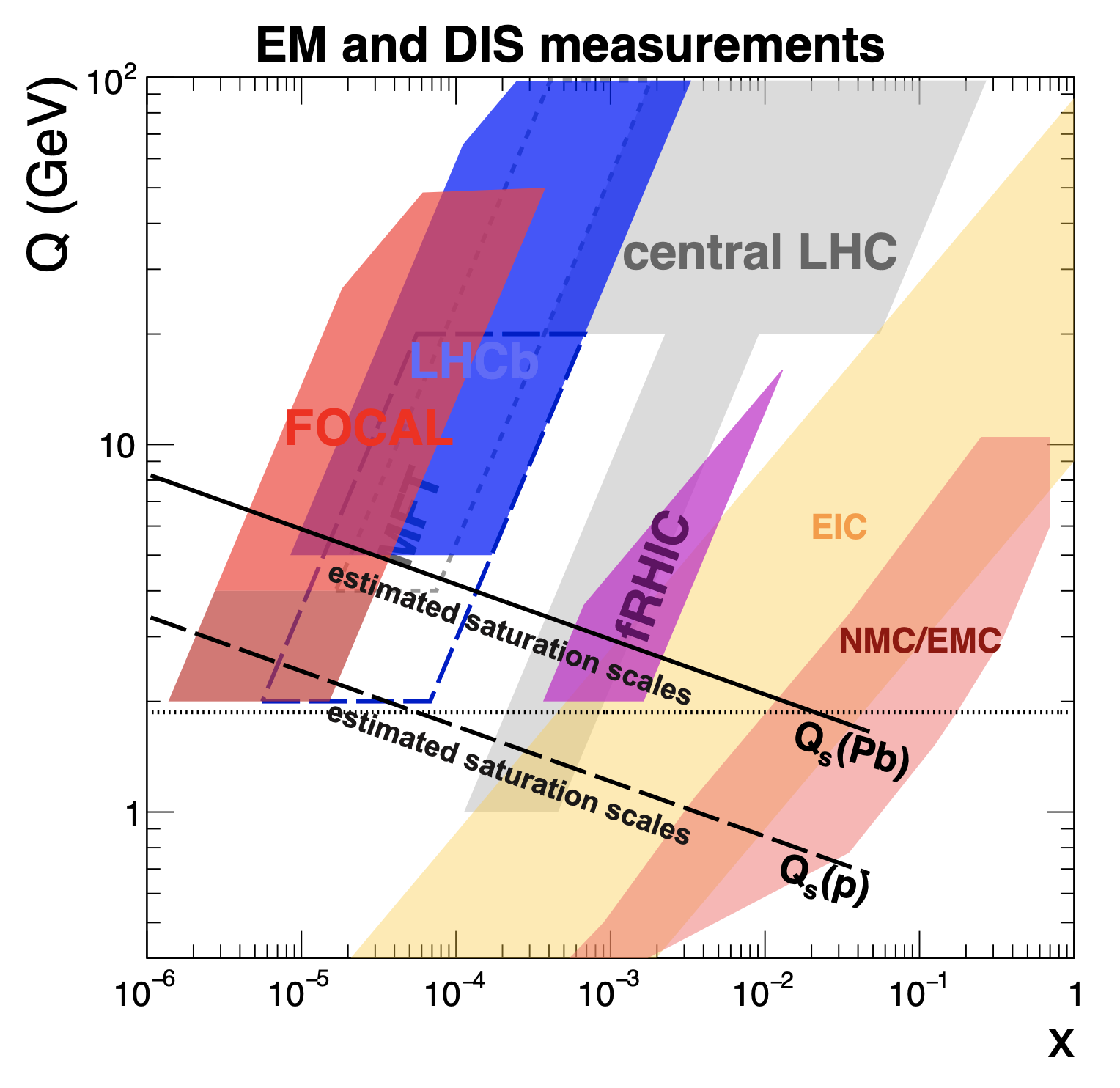}}
\caption{Approximate (x,Q) coverage for measurement of deep inelastic scattering in various experiments.}
\label{FigF}
\end{figure}

The Forward Calorimeter project (FoCal) extends the scope of ALICE by adding new capabilities to explore the small-x parton structure of nucleons and nuclei. In particular, the FoCal provides unique capabilities at the LHC to investigate Parton Distribution Functions in the as-yet unexplored regime of Bjorken-x down to \mbox{x  $\sim$ 10$^{-6}$} and low momentum transfer \mbox{Q $\sim$ 4 GeV/c} (Fig.~\ref{FigF}), where it is expected that the hadronic structure evolves non-linearly due to the high gluon densities. Such effects are a necessary consequence of the non-Abelian nature of quantum chromodynamics (QCD), and their observation and characterisation would be a landmark in our understanding of the strong interaction. The main goals of the FoCal physics program are to: (i) quantify the nuclear modification of the gluon density in nuclei at small-x and Q$^{2}$ by measuring isolated photons in proton--proton and proton--lead collisions; (ii) investigate non-linear QCD evolution by measuring azimuthal $\pi^{0}$--$\pi^{0}$ correlations and isolated $\gamma$--$\pi^{0}$ correlations in proton--proton and proton--lead collisions; (iii) investigate the origin of long range flow-like correlations by correlating neutral meson production over a large range in rapidity in proton--proton and proton--lead collisions; (iv) quantify parton energy loss at forward rapidity by measuring high-$p_{\rm T}$ neutral pion production in lead--lead collisions.

\begin{figure}[tb]
\centerline{\includegraphics[width=12.5cm]{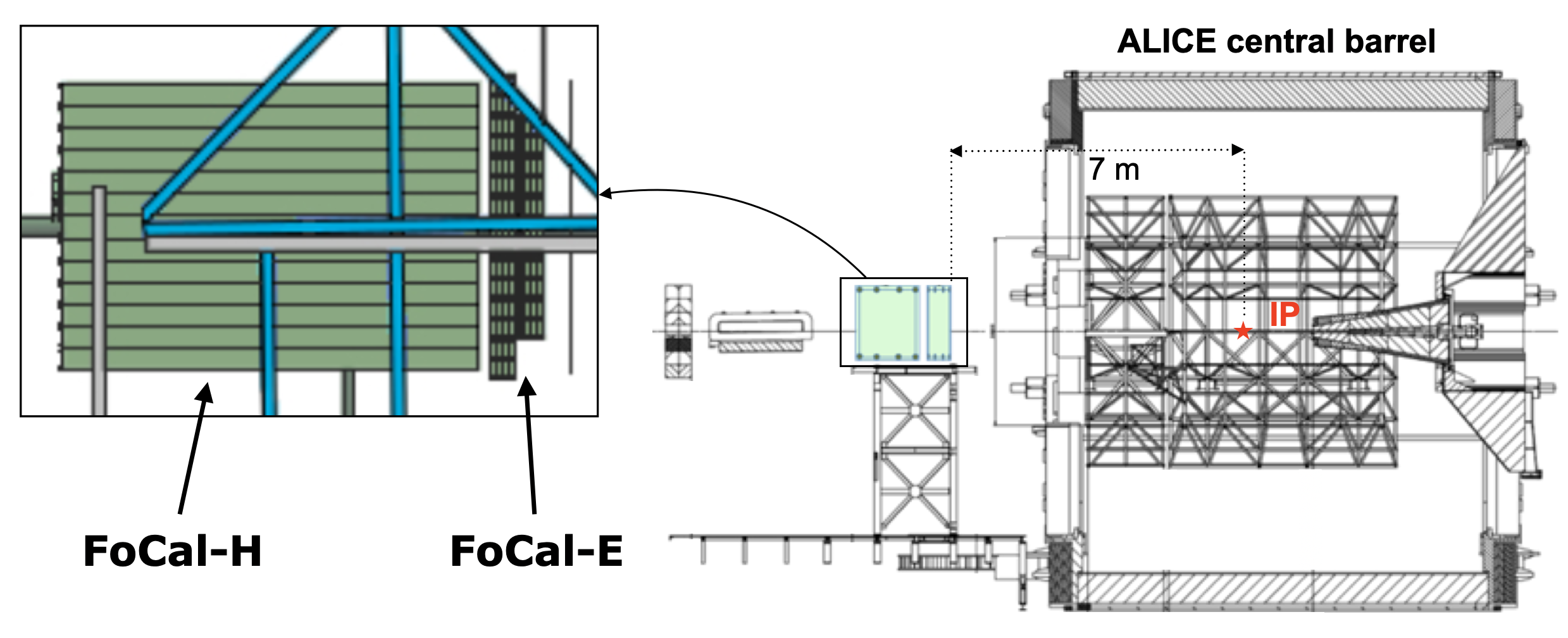}}
\caption{ Installation of the FoCal at the 7 m location from the interaction point (IP) with FoCal-E and FoCal-H detectors.}
\label{FigG}
\end{figure}

The FoCal design is presented in Fig.~\ref{FigG}. The detector will be placed outside the ALICE solenoid magnet at 7~m from the interaction point, covering pseudo-rapidity range 3.4 $< \eta <$ 5.8, and will consist of an electromagnetic (FoCal-E) and a hadronic (FoCal-H) calorimeter. 

\begin{figure}[tb]
\centerline{\includegraphics[width=12.5cm]{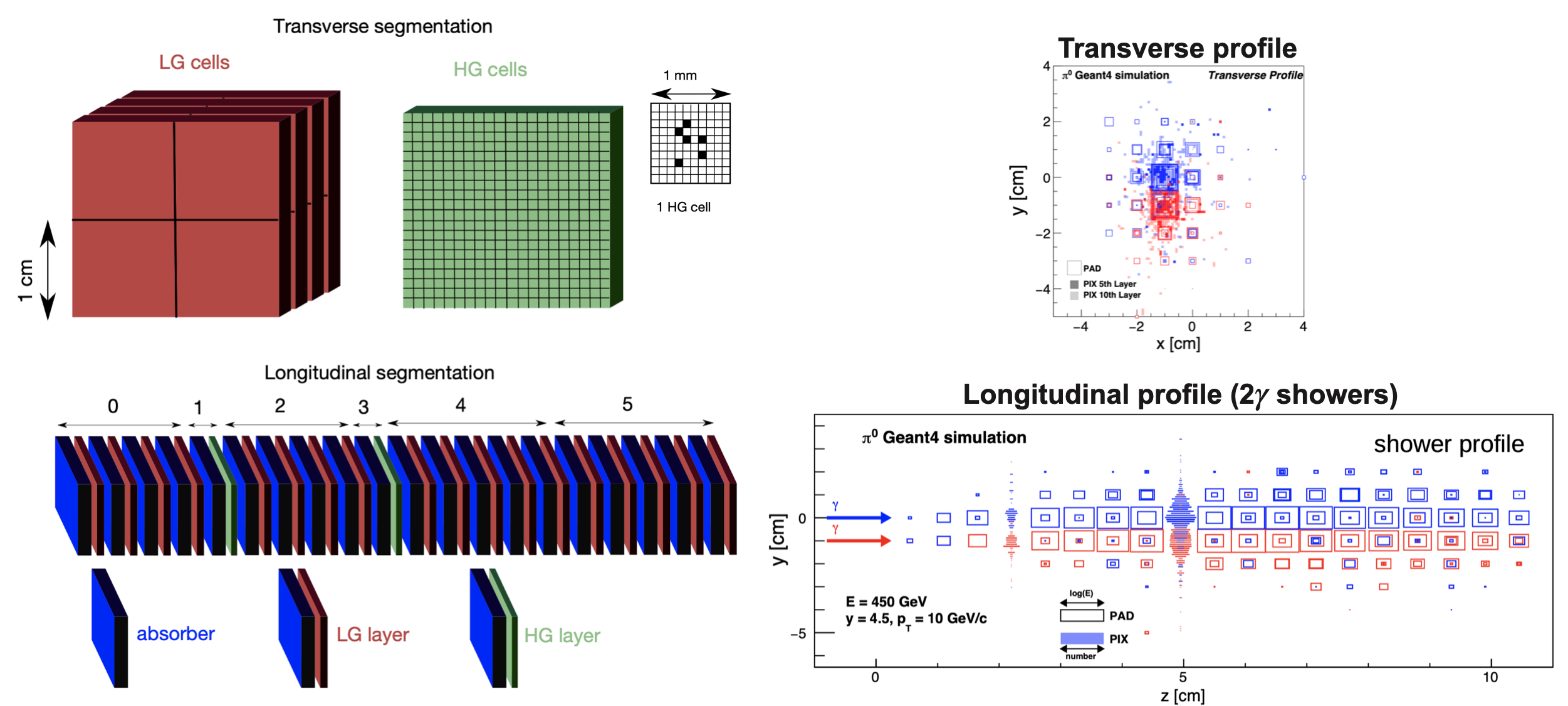}}
\caption{(Left) Schematic view of the structure of the FoCal-E detector. (Right top) Longitudinal and transverse profile of two showers produced in the FoCal-E detector by two photons.}
\label{FigH}
\end{figure}

\begin{figure}[tb]
\centerline{\includegraphics[width=12.5cm]{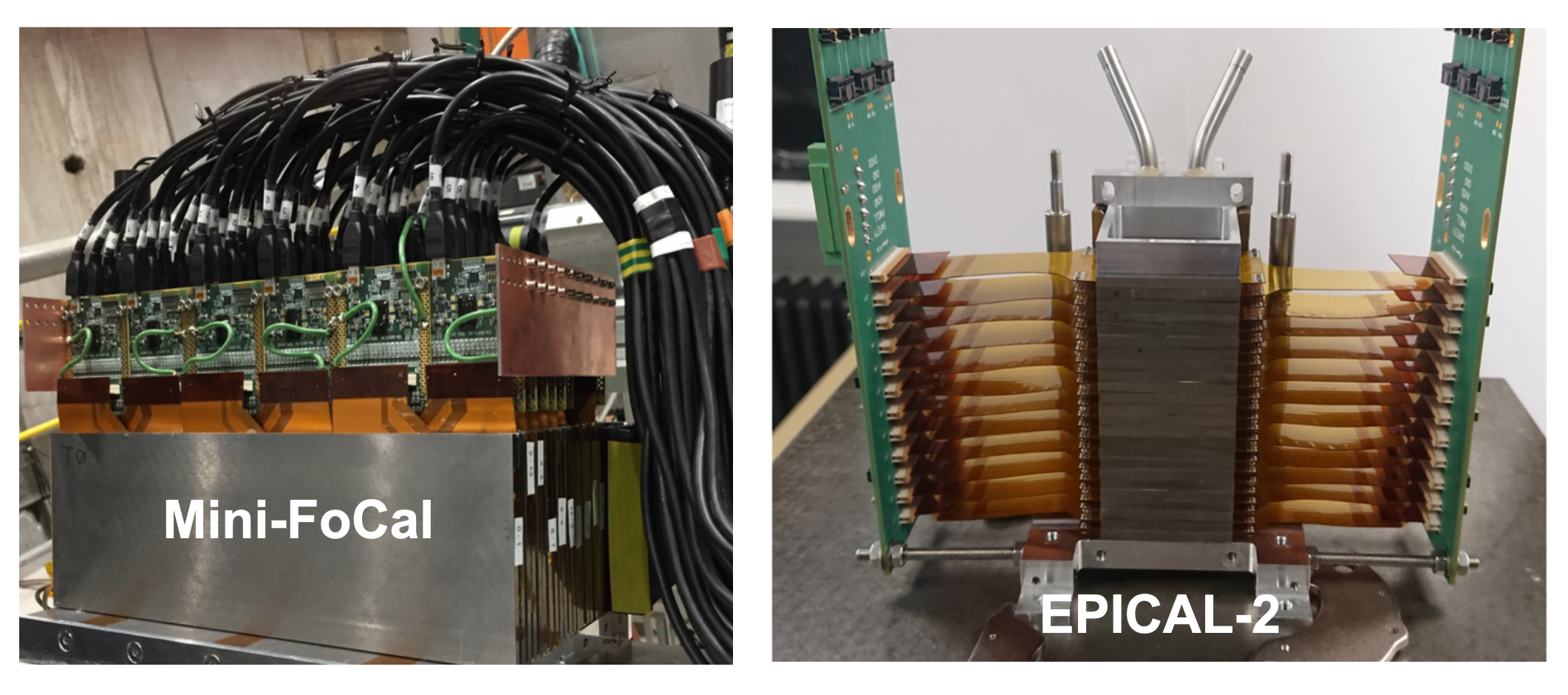}}
\caption{Pictures of the Mini-FoCal (left) and EPICAL-2 (right) prototypes.}
\label{FigI}
\end{figure}

To separate $\gamma$ and $\pi^{0}$ at high energy it is necessary to minimise occupancy effects and optimise photon shower separation in the FoCal-E. Tungsten is the best choice for the absorber material, having small Molière radius (R$_{M}$ = 9~mm) and radiation length (X$_{0}$ = 3.5~mm). A fine transverse granularity readout is granted using silicon pixel sensors. The detector design foresees a 20 X$_{0}$ thick silicon-tungsten sampling calorimeter with 18 layers of tungsten absorbers and silicon sensors with two different granularities (Fig.~\ref{FigH}, left). Low-granularity silicon sensors (pad size 1 cm$^{2}$) with fast integration time for charge collection will be used in 16 out of the 18 layers and will provide shower profile and total energy measurement. The 2 remaining layers shall be equipped with high-granularity monolithic active pixel sensors, like the ALPIDE sensor, characterised by slower integration time (about 5 $\mu$s). The high-granularity layers will be placed at the positions, where an electromagnetic shower reaches its maximum, to improve two-shower separation. A total sensor area of 14.5~m$^{2}$ and 1.5~m$^{2}$ are expected for the low- and high-granularity layers, respectively, with a total of about 150$\cdot10^{3}$ individual pad channels and about 4$\cdot10^{3}$ pixel sensors. Two examples of prototypes assembled as part of the FoCal-E R\&D project can be seen in Fig.~\ref{FigI}. 
Mini-FoCal is an assembly of 20 layers, each consisting of a 3.5~mm thick tungsten plate followed by a 0.3~mm thick silicon sensor. These are Hamamatsu silicon pad sensors having an 8 $\times$ 8 pad matrix, with each pad 1~cm$^{2}$ large. A first version of this prototype was tested at PS and SPS showing good linearity and energy resolution $\sigma_{e}/E$ = $0.17/\sqrt{E} \bigoplus 0.019$\cite{FoCalLoI}. A second version (shown in Fig.~\ref{FigI}, left) was placed in the ALICE cavern during the proton--proton 13~TeV physics data taking in 2018 to measure the background. 
EPICAL, is a small fully digital silicon-tungsten calorimeter with high granularity, based on MIMOSA sensors in its first version\cite{EPICAL} and on ALPIDE sensors in its second assembly. EPICAL-2 (shown in Fig.~\ref{FigI}, right) has 24 layers consisting of a 3~mm thick tungsten absorber followed by 2 ALPIDE sensors, covering a transverse surface of 3~$\times$~3~cm$^{2}$. It was tested with beams at DESY during 2019-2020, showing that MAPSs are suitable for such application.

FoCal-H is needed for photon isolation and jet measurements. It can be built as a conventional sampling hadronic calorimeter with a thickness of $\sim$6 hadron interaction lengths and a total length in z direction of $\sim$1.1~m. The transverse size will be similar to that of FoCal-E. Present design foresees a total of $\sim$1000 scintillating fiber based towers with transverse dimension in the range 2 -- 5~cm. A copper radiator prototype has been assembled and is going under test beam during the 2021. The readout will be based on avalanche photodiodes (APDs) or silicon photomultipliers (SiPMs).

%_____________________________________
\section{ALICE~3: prospectives after LHC Run~4}
The ALICE Collaboration is preparing a Letter of Intent for the construction of a new LHC experiment, to be submitted to the LHC Experiments Committee (LHCC) in 2022. The goal is to extend the LHC heavy-ion program beyond Run~4 with a dedicated experiment that provides much improved vertex resolution, larger rapidity coverage and clean electron identification combined with high rate capabilities \cite{ALICE3EoI}. In particular it should be able to measure the production of leptons, photons and identified hadrons down to $p_{\rm T}$ scales of the order of a few tens of MeV/c. Examples of topics that gain significant advances are:

\begin{itemize}
\item Hadronization mechanism from QGP via multi-charm hadrons measurement.
\item Microscopic description of the dynamics of heavy quarks in a QGP via bound state (quarkonia and exotica) formation and dissociation mechanisms.
\item The description of the early phases of the collision through real and virtual photons measurement.
\item Experimental evidence for the restoration of chiral symmetry in the hot and dense phase via precision measurement of the thermal dilepton continuum starting from the $\rho$ meson and reaching up to masses of about 1.6~GeV.
\item Low’s theorem verification through photon measurement at very low $p_{\rm T}$, below 10~MeV/c.
\end{itemize}

\begin{figure}[tb]
\centerline{\includegraphics[width=9cm]{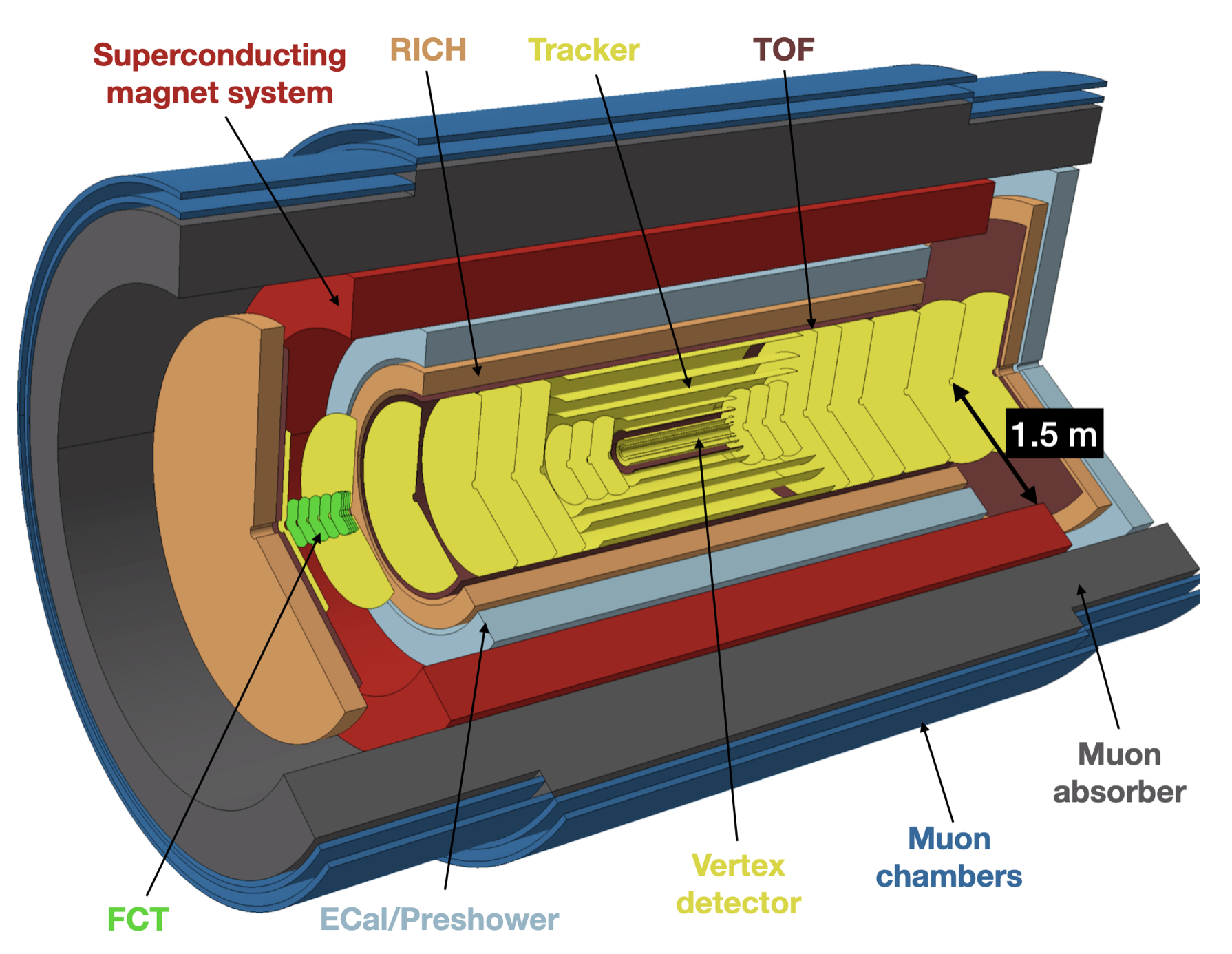}}
\caption{Schematic view of the ALICE~3 detector proposal.}
\label{FigL}
\end{figure}

The observables mentioned in the previous list determine the requirements on the detector design;  a schematic view is available in Fig. \ref{FigL}. The detector, which covers the pseudorapidity region of $|\eta|$~$<$~4 over the full azimuth, has a very compact layout. Main tracking system and PID detectors, assembled as a central barrel and two end-caps, are surrounded by a superconducting magnet system with internal radius of 1.5~m and longitudinal dimension of 4~m. Externally to the magnet there are only the muon chambers, preceded by the muon absorber. 

To provide the required distance of closest approach (DCA) resolution to the primary vertex, a three layer ultra-light bent silicon detector is expected to go within the beam pipe (inside a secondary vacuum). A retractable design is considered to provide the required beam aperture during beam injection, with 5~mm minimum radial distance to the interaction point for the innermost layer. The possibility to reach the appropriate spatial resolution and material budget are explored in the R\&D activities within the ITS~3 project with the application of the 65~nm CMOS technology and the possibility to bend large dimension thin silicon sensors with removal of mechanical support and cooling infrastructures. R\&D challenges concern mechanical supports, cooling and radiation tolerance.

Tracking will be completed with more detector layers equipped with silicon MAPS sensors arranged in modules with water-cooling and carbon-fiber space frame for mechanical support. In this case, the requirements on spatial resolution and material budget are less stringent and the appropriate design of the magnetic field is essential to obtain the needed transverse momentum resolution. The total silicon surface is expected to be $\sim$60~m$^2$ so it is crucial to develop cost-effective sensor and to automatise as much as possible the module production. 

Hadron heavy-flavour decay reconstruction requires $\pi$/K/p separation for transverse momentum up to a few GeV/c. This will be provided with a Time Of Flight (TOF) detector, complemented with a Cherenkov (RICH) detector for higher particle momentum. TOF detector will consist of one barrel layer plus two end-caps (one on each side) made with silicon sensor characterised by timing resolution of $\sim$20~ps. At least three sensor technologies are under consideration for this application: Low Gain Avalanche Diodes (LGAD), MAPS and Single Photon Avalanche Diode (SPAD). Total silicon surface in this case is $\sim$45~m$^2$ requiring cost-effective sensor development.     
In the Cherenkov detector, using aerogel as radiator material, the sensitive $p_{\rm T}$ window can be matched to the one of TOF in order to ensure continuity in the PID capabilities. Present strategy foresees photons collection based on SiPMs; an alternative based on MAPS is also considered. End-cap layers are foreseen also for this detector. TOF and RICH are useful for electron identification; a large pion rejection up to few GeV/c is required for dielectrons and quarkonia measurements. 

Muon chambers are foreseen as the outermost layers of the detector in the central barrel region. This system is fundamental for quarkonia reconstruction and in particular for charmonia (J/$\Psi$) down to $p_{\rm T} = 0$, complementing the electron identification capabilities. Baseline detector technology is Resistive Plate Chamber (RPC), but other options will also be considered.

Photons and jets measurements require a large acceptance electromagnetic calorimeter. This is expected to be placed between RICH detector and magnet cryostat, covering a large pseudorapidity range in the central region plus a layer in the end-cap. 

Ultra-soft photon measurement requires a dedicated detector in the forward region, namely the Forward Conversion Tracker (FCT), based on silicon pixel sensors. Basic requirement is the minimisation of material budget in front of the detector.

%_______________
\section{Conclusions}	
ALICE Collaboration successfully completed the upgrade program and is ready to start to accomplish the reach scientific program foreseen for LHC Run~3. Intense R\&D activities are ongoing to develop the near future upgrades, including installation of two new detectors to supplement the physics goals during LHC Run~4. A proposal for the construction of a new LHC experiment to continue the heavy-ion program after LHC Run~4 and push forward our knowledge of the QCD behaviour in dense and hot conditions has been made by the Collaboration and a Letter of Intent is in preparation.

%_________________________________
%\begin{thebibliography}{000} %for 3 digits
%\begin{thebibliography}{00}  %for 2 digits

\end{document}